\documentclass{article}
\usepackage{spconf_asru,amsmath,graphicx}

\usepackage{xcolor,soul,framed} 

\colorlet{shadecolor}{yellow}

\usepackage{array}
\usepackage{url}

\usepackage{cite}
\usepackage{booktabs} 
\usepackage{multirow}
\usepackage{multicol}
\usepackage{amsfonts}
\usepackage{float}
\usepackage{bbold}

\usepackage{hyperref}
\usepackage{etoolbox,siunitx}
\robustify\bfseries
\usepackage{balance}

\usepackage{amssymb}
\usepackage{pifont}
\newcommand{\cmark}{\ding{51}}%
\newcommand{\xmark}{\ding{55}}%

\DeclareMathOperator*{\argmin}{arg\,min}  
\DeclareMathOperator*{\argmax}{arg\,max}  

\usepackage{scalerel}

\title{Late Audio-Visual Fusion for In-The-Wild Speaker Diarization}
\name{Zexu Pan$^{1,2}$, Gordon Wichern$^1$, Fran\c{c}ois G. Germain$^1$, Aswin Subramanian$^1$, Jonathan Le Roux$^1$
\thanks{This work was performed while Zexu Pan was an intern at MERL.}
}
\address{
  $^1$Mitsubishi Electric Research Laboratories (MERL), Cambridge, MA, USA\\
  $^2$Institute of Data Science, National University of Singapore, Singapore}
\begin{document}
\ninept
\maketitle

\setlength{\abovedisplayskip}{4pt}
\setlength{\belowdisplayskip}{4pt}

\begin{abstract}

Speaker diarization is well studied for constrained audios but little explored for challenging in-the-wild videos, which have more speakers, shorter utterances, and inconsistent on-screen speakers. We address this gap by proposing an audio-visual diarization model which combines audio-only and visual-centric sub-systems via late fusion. For audio, we show that an attractor-based end-to-end system (EEND-EDA) performs remarkably well when trained with our proposed recipe of a simulated proxy dataset, and propose an improved version, EEND-EDA++, that uses attention in decoding and a speaker recognition loss during training to better handle the larger number of speakers. The visual-centric sub-system leverages facial attributes and lip-audio synchrony for identity and speech activity estimation of on-screen speakers. Both sub-systems surpass the state of the art (SOTA) by a large margin, with the fused audio-visual system achieving a new SOTA on the AVA-AVD benchmark.

\end{abstract}

\section{INTRODUCTION}
\vspace{-.2cm}
\label{sec:intro}

As speech signals are rich in information such as emotion, identity, and location, speech processing algorithms have been proposed for many tasks such as emotion recognition, speaker identification, and localization~\cite{pan2020multi, snyder2018x, qian2021multi, pan2022hybrid}. However, these algorithms are often optimized for isolated speech segments. Therefore, an effective preprocessing algorithm to find ``who spoke when'', i.e., speaker diarization~\cite{landini2022bayesian}, is highly desirable. Classical examples of speaker diarization algorithms~\cite{park2022review,landini2022bayesian} are cascaded systems with multiple stages: voice activity detection (VAD), frame segmentation, speaker embedding extraction, and clustering. Each stage is optimized independently, as a result, errors can accumulate and significantly impact the overall performance. Additionally, classical algorithms often cannot handle overlapping speakers as each frame is assigned to a single speaker.

In recent years, audio-only end-to-end (E2E) neural-network-based speaker diarization algorithms have gained prominence, offering simplified model-building and inference processes while often achieving better performance. These algorithms also naturally handle speaker overlap by decoding a speech activity stream for each speaker. One of those systems is end-to-end neural diarization (EEND) with an encoder-decoder based attractor (EDA) calculation module (EEND-EDA)~\cite{horiguchi2020end, horiguchi2022encoder}. EEND-EDA has gained increasing attention due to its ability to flexibly handle an unknown number of speakers, using an iterative process that extracts individual speakers until a stop flag is raised indicating that all speakers have been estimated, in contrast to methods relying on a hard-coded number of streams~\cite{fujita2019end}. In both cases, permutation invariant training (PIT)~\cite{Isik2016Interspeech09, yu2017permutation} is required to address the speaker order ambiguity between the network outputs and the ground-truth labels.

Audio-visual methods have also been explored for speech understanding, inspired by the multimodal aspect of human perception~\cite{smith2005development}. Indeed, we often rely on various sensory stimuli for understanding and attending to speakers in a conversation, such as observing facial expressions~\cite{ephrat2018looking}, lip movements~\cite{usev21}, and body gestures~\cite{pan2022seg}. Several existing audio-visual diarization algorithms leverage such synergies between speech signals and visual features. The WST model~\cite{chung2019said} enrolls speakers based on audio-visual correspondence in a cascaded diarization system. E2E methods~\cite{qiu2022visual,he22c_interspeech} fuse audio and visual representations, training similarly to audio-only EEND~\cite{fujita2019end}.

\begin{figure}[t]
\begin{minipage}[t]{.49\linewidth}
  \centering
  \centerline{\includegraphics[width=\linewidth]{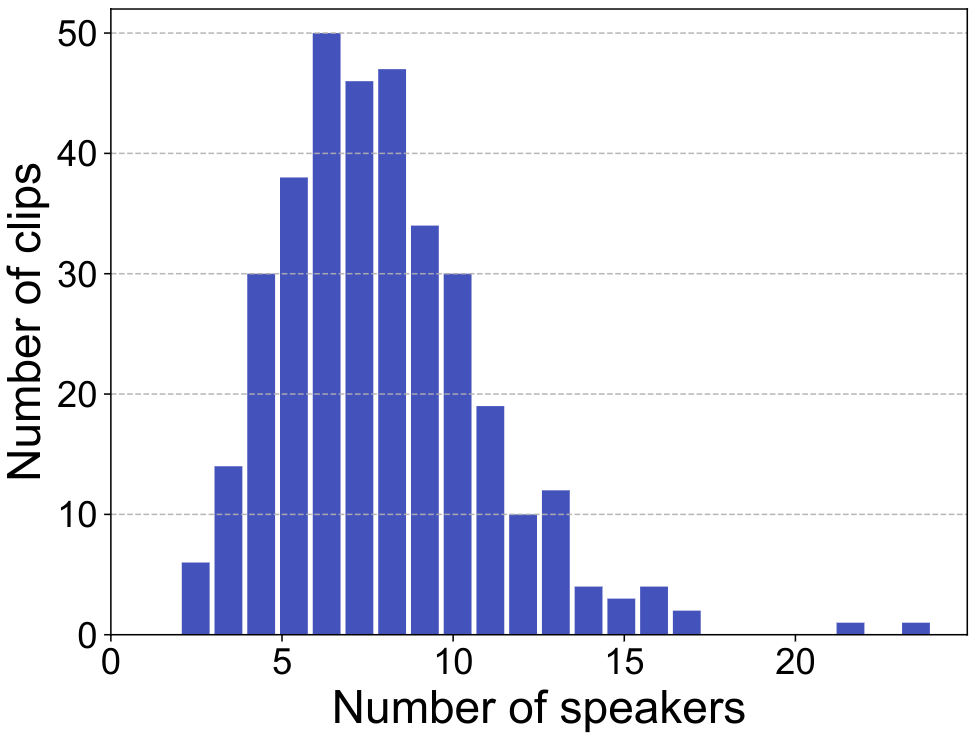}}
  \vspace*{-1mm}
\end{minipage}
\hfill
\begin{minipage}[t]{.49\linewidth}
  \centering
  \centerline{\includegraphics[width=\linewidth]{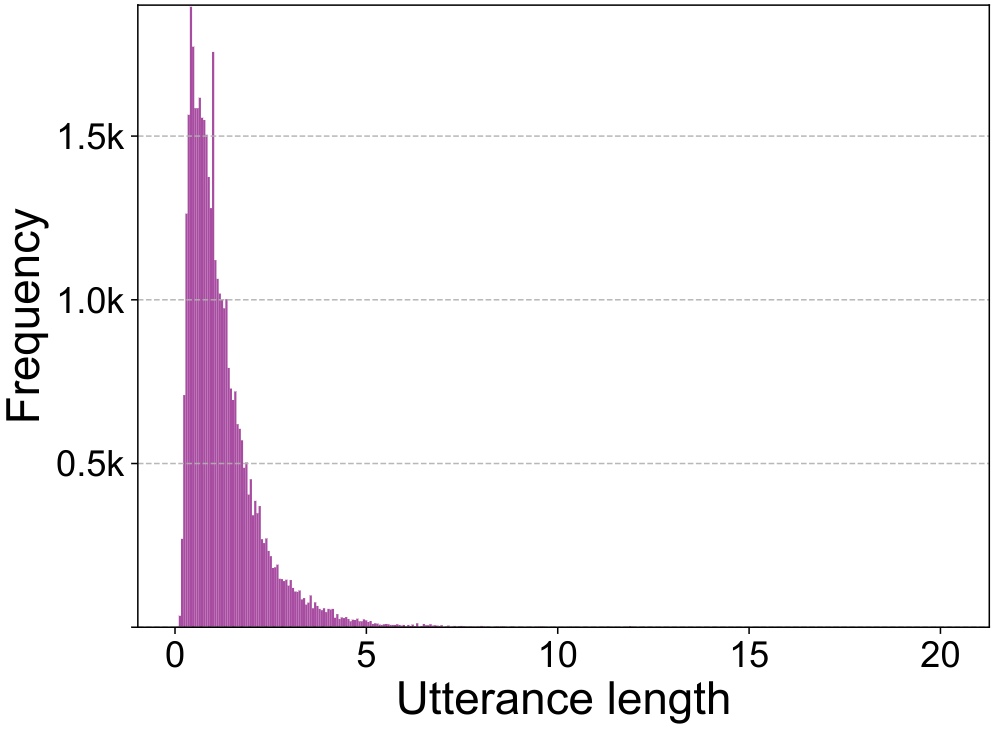}}
  \vspace*{-1mm}
\end{minipage}
\vspace*{-3mm}
\caption{Histograms of the number of speakers (left) and the utterance length in seconds (right) of the audio recordings in the AVA-AVD dataset, plots are taken from~\cite{xu2021ava}.}
\label{fig:dataset}
\vspace*{-6mm}
\end{figure}

\begin{figure*}[t]
  \centering
  \includegraphics[width=0.99\linewidth]{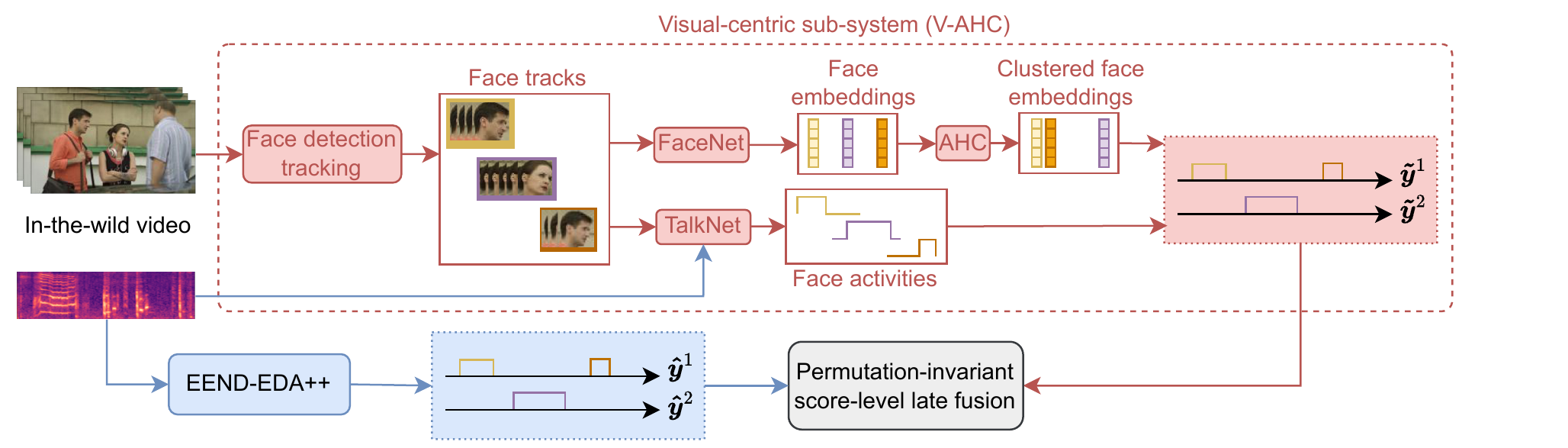}
  \vspace*{-4mm}
  \caption{Our proposed audio-visual speaker diarization model named AV-EEND-EDA++. The blue path is the audio-only sub-system named EEND-EDA++, while the red path is the visual-centric clustering-based sub-system named V-AHC. The diarization results of the EEND-EDA++ and V-AHC sub-systems are fused with a permutation-invariant score-level late fusion. The figure is best viewed in color.}
  \vspace*{-2mm}
\label{fig:av_model}
\end{figure*}

\begin{figure}[t]
  \centering
  \includegraphics[width=\linewidth]{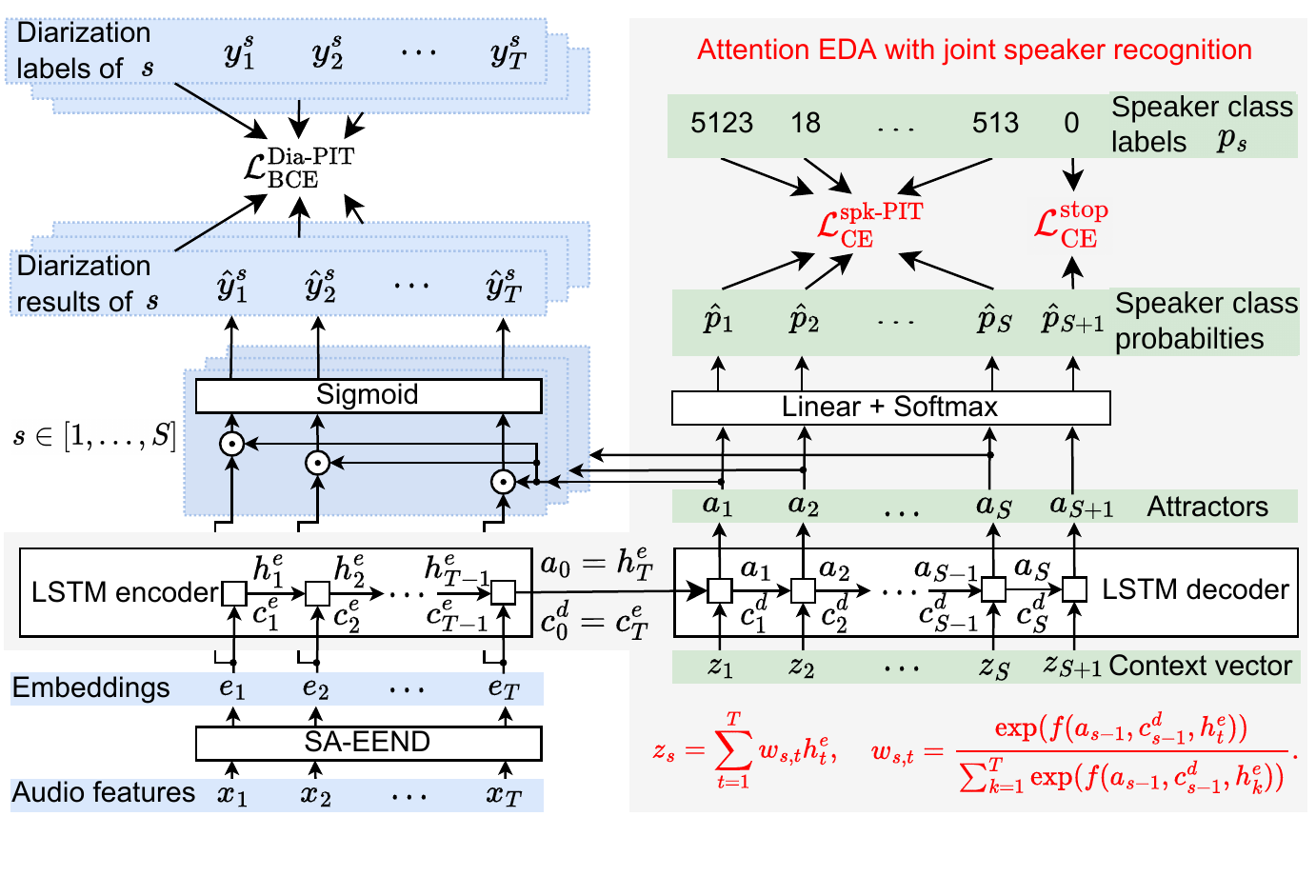}
  \vspace*{-10mm}
  \caption{Our proposed EEND-EDA++. We introduce an attention mechanism in EDA to increase its capacity in decoding more speaker attractors, and propose to train the attractors on speaker recognition to improve the attractor representation. The symbol $\odot$ refers to the inner product. Novel contributions are marked in red.}
  \vspace*{-4mm}
\label{fig:audio_network}
\end{figure}

The audio-only EEND-EDA model and the audio-visual diarization models have demonstrated remarkable performance in scenarios involving meetings or conversations~\cite{ryant2020third}, where the speaker count is typically limited to around 10 people and all speakers are on-screen. However, it is untested in situations such as movies or in-the-wild recordings, where a larger number of speakers may be present, and the duration of many utterances can be significantly shorter, as exemplified by the statistics of the recent AVA-AVD dataset~\cite{xu2021ava} (see Fig.~\ref{fig:dataset}). In that dataset, a 5-minute movie clip can contain up to 20 speakers, with much shorter typical individual utterances compared to meetings, which especially test the robustness of PIT with its factorial complexity with increasing number of speakers, and the capacity of the EDA module to identify representative attractors, considering the increased complexity of distinguishing and representing a larger set of speakers. Additionally, recordings contain a wide variety of background noises, sound effects, and music.

In the literature, the AVR-Net~\cite{xu2021ava} and DyViSE~\cite{dyvise2022} target the challenges of such in-the-wild videos, encompassing more speakers, including off-screen and irrelevant speakers, and complex acoustic scenes. These approaches enhance the speaker embedding extraction stage through audio-visual early fusion in a classical cascaded diarization algorithm.
Alternatively, we advocate for late fusion that allows the visual signals to play at least two important roles that could help diarization: the synchronization between the lip movements and the speech signals provides strong hints about the speech activity~\cite{tao2021someone}, and the facial attributes provide robust evidence about speaker identity~\cite{serengil2020lightface}.

In this work, we aim to build an audio-visual speaker diarization system targeting in-the-wild videos. We propose a model named AV-EEND-EDA++, which is depicted in Fig.~\ref{fig:av_model}. It comprises an audio-only sub-system named EEND-EDA++, shown in Fig.~\ref{fig:audio_network}, that is built upon the existing EEND-EDA system for in-the-wild audio recording, and a visual-centric sub-system named V-AHC, shown in the top part of Fig.~\ref{fig:av_model}, that performs on-screen speaker diarization based on available on-screen face tracks. The two sub-systems are combined with a permutation-invariant late-fusion technique based on speaker activity probability.

For the audio-only sub-system, we first investigate the capability of the EEND-EDA system on in-the-wild movie-style audio recordings from the AVA-AVD benchmark~\cite{xu2021ava}. 
In particular, we provide here a recipe such that the simulated dataset on which the system is pre-trained ends up matching the distribution of movie-style audio recordings.
We demonstrate that the EEND-EDA system trained on our simulated dataset surpasses the performance of classical cascaded systems on the AVA-AVD benchmark. For EEND-EDA++, we propose an attention-based EDA module to enhance the network's capacity when decoding a large number of speaker attractors, which allows the system to better focus on relevant information and improves the overall diarization performance. We also propose a joint training approach that trains the speaker attractors on a speaker recognition task to enhance the discriminative power of the attractor representation.

While the EEND-EDA++ sub-system focuses on audio-only speaker diarization, considering the robustness of visual signals against acoustic noise, 
we propose the visual-centric sub-system V-AHC that explicitly leverages visual signals to enhance the speech activity detection and model the speaker identity for on-screen speaker diarization. The V-AHC model utilizes the active-speaker detection algorithm TalkNet~\cite{tao2021someone} to determine on-screen speaker activity and speaker changes, and employs agglomerative hierarchical clustering (AHC) based on face embeddings extracted from FaceNet~\cite{schroff2015facenet} for speaker clustering. 

Experimental results show that the speaker identity of our EEND-EDA++ is better preserved across recordings as a result of our proposed speaker recognition loss. Both our proposed audio-only sub-system EEND-EDA++ and visual-centric sub-system V-AHC outperform existing audio-visual diarization algorithms on the AVA-AVD diarization benchmark~\cite{xu2021ava}, with our audio-visual late-fusion system AV-EEND-EDA++ achieving the new state of the art in terms of diarization error rate and Jaccard error rate.

\section{PROPOSED AV-EEND-EDA++}
Our proposed AV-EEND-EDA++ consists of the audio-only sub-system EEND-EDA++ and the visual-centric sub-system V-AHC. In this section, we present each of the sub-systems first, followed by the permutation-invariant late-fusion strategy.

\vspace{-.2cm}
\subsection{Proposed audio-only EEND-EDA++}
\label{sec:methodology}

\vspace{-.1cm}
\subsubsection{Related work: EEND-EDA}
\label{sec:related_works}
\vspace{-.1cm}

EEND-EDA~\cite{horiguchi2020end} is an E2E diarization model that can handle a flexible number of speakers as well as overlapping speakers. Given a recording, it encodes the audio features into a sequence of audio embeddings $e_t\,{\in}\,\mathbb{R}^{D}$, $t\,{\in}\,[1, \dots, T]$, using stacked Transformer encoders~\cite{vaswani2017attention} without positional encoding, where $D$ is the embedding dimension and $T$ the total number of audio frames. The encoder is denoted as \mbox{SA-EEND}, for self-attentive EEND. Next, in the EDA module, an LSTM encoder encodes the time-shuffled $e_t$. Based on the last hidden and cell states of the encoder, an LSTM decoder estimates a flexible number of attractors $a_s\,{\in}\,\mathbb{R}^D$ ($s\,{\in}\,[1,\dots, S]$) that represent speakers. The estimation process is controlled by a stop flag, representing the complete estimation of all speakers. Speaker activity at time $t$ for each speaker is obtained by taking the inner product of each attractor with the audio embedding $e_t$ and applying a sigmoid.

In follow-up work~\cite{horiguchi2021towards}, the authors found that the number of output speakers is capped by the maximum found in the pre-training dataset, even if the model is adapted to other datasets with more speakers, due to the EDA operation. A direct solution is to pre-train the model on a dataset with more speakers. However, the EDA decoder still needs to estimate many attractors based only on the last hidden and cell states of the EDA encoder. Furthermore, the attractors are only trained to encode speaker information that correlates with the audio embeddings of the frames where the corresponding speakers are active, without a mechanism to ensure this information remains valid across recordings.
We propose to extend EEND-EDA to overcome both these limitations.

\vspace{-.2cm}
\subsubsection{Attention EDA}
\vspace{-.1cm}
The original EDA module iteratively estimates an attractor for every speaker in the recording. At each iteration $s$, the EDA LSTM decoder takes as input the previous attractor $a_{s-1}$, the previous cell state $c^d_{s-1}$, and a zero vector, resulting in a limited capacity to memorize all speakers, and limiting performance when the number of speakers in a recording is large.

Inspired by the attention in machine translation~\cite{bahdanau2014neural}, we propose to use an attention EDA LSTM decoder here, such that the estimation of attractor $a_s$ is conditioned on a distinct context vector $z_s$ instead of the zero vector, with $z_s$ computed as a weighted sum of the EDA encoder outputs $h^e_t$: 
\begin{equation}
    z_s = \sum_{t=1}^T w_{s,t}h^e_t,
\end{equation}
where
\begin{equation}
    w_{s,t} = \frac{\exp(f(a_{s-1},c^d_{s-1},h^e_{t}))}{\sum_{\tau=1}^T \exp(f(a_{s-1},c^d_{s-1},h^e_{\tau}))}.
\end{equation}
We parameterize $f(\cdot)$ as a one-layer feedforward neural network with hyperbolic tangent (tanh) activation. We denote this method EEND-EDA+Att. It is worth noting that, in contrast with the original EEND-EDA, we use the SA-EEND with positional encoding, and we do not need to shuffle the audio embeddings $e_t$ before passing them to the EDA LSTM encoder as EEND-EDA did.

\vspace{-.2cm}
\subsubsection{Objective functions for speaker recognition}
\vspace{-.1cm}
Besides the diarization loss, the EEND-EDA attractors are trained via a binary classification loss in which the attractors iteratively predict whether there are remaining unaccounted-for speakers in the recording, or there are no more speakers. The attractors are thus not explicitly trained to encode speaker information that remains valid across recordings, such as speaker identity. Since each speaker's activity is conditioned on its attractor, it is expected that an attractor better correlated with speaker identity will benefit the diarization task. Inspired by speaker extraction approaches in which the extraction is conditioned on an attractor that is jointly trained to recognize the speaker~\cite{vzmolikova2017learning,pan2020muse,pan2021reentry}, we also train the attractors using a speaker recognition loss, such that the attractors explicitly represent speaker information. We refer to this approach as EEND-EDA+Spk.

We employ a softmax layer to transform each attractor $a_s$ into a probability distribution $\hat{p}_s$ over the speakers in the dataset. During training, we define the ground-truth speaker class labels $p_s$, for $s\,{\in}\,[1,\dots, S],$ using the actual number of speakers $S$ in a recording, where $p_s(j)$, for $j\,{\in}\, [1,\dots, J],$ is a binary label indicating if the $s$-th speaker in the recording is the $j$-th speaker in the speaker dataset, and $J$ denotes the total number of speakers in the training set. We introduce an additional class representing ``Not a speaker,'' corresponding to $j\!=\!0$, and use it as a stop flag, such that the network learns to stop decoding attractors at inference time whenever an attractor falls into this class. The speaker classification objective function for the first $S$ attractors is defined as:
\begin{equation}
    \mathcal{L}_{\text{CE}}^{\text{spk-PIT}} = \argmin_{\pi \in \mathcal{P}_S}  -\!\sum_{s=1}^{S} \sum_{j=0}^{J} p_{\pi(s)}(j) \log \hat{p}_s(j),
\end{equation}
where $\mathcal{P}_S$ denotes the set of permutations over $\{1,\dots,S\}$. We use PIT to find the optimum permutation order between the estimated attractors and the speaker labels, relying on Sinkhorn's algorithm (SinkPIT)~\cite{tachibana2021towards} to avoid the factorial complexity against the number of speakers. 
At the same time, the $S\!\!+\!\!1$-th attractor is trained to fall into the ``Not a speaker'' class, using the objective function defined as:
\begin{equation}
    \mathcal{L}_{\text{CE}}^{\text{stop}} = - \sum_{j=0}^{J} p_{S+1}(j) \log \hat{p}_{S+1}(j) = - \log \hat{p}_{S+1}(0).
\end{equation}

\vspace{-.2cm}
\subsubsection{Overall objective function}
\vspace{-.15cm}

Our final proposed audio-only sub-system combines EEND-EDA+Att and EEND-EDA+Spk, and is referred to as EEND-EDA++ shown in Fig.~\ref{fig:audio_network}. The overall objective function is defined as:
\begin{equation}
\mathcal{L}_{\text{all}} = \mathcal{L}_{\text{BCE}}^{\text{Dia-PIT}} + \beta (\mathcal{L}_{\text{CE}}^{\text{spk-PIT}} + \alpha \mathcal{L}_{\text{CE}}^{\text{stop}}),
\end{equation}
where
\begin{equation}     
\resizebox{\columnwidth}{!}{%
$\displaystyle{\mathcal{L}_{\text{BCE}}^{\text{Dia-PIT}}\! =\! \displaystyle \argmin_{\pi \in \mathcal{P}_S} -\!\! \displaystyle \sum_{s=1}^S \sum_{t=1}^T ( \gamma y^{\pi(s)}_t \!\log (\hat{y}_t)\! +\! (1\!-\!y^{\pi(s)}_t)\!\log (1-\hat{y}_t)\!)}$}
\end{equation}
is the objective of the diarization task, with $y_t$ and $\hat{y}_t$ respectively denoting the ground-truth and estimated speaker activity probabilities at frame $t$. We again use the SinkPIT algorithm to reduce the computational complexity of increasing speakers. Different loss terms are balanced using scalar weights $\alpha$ and $\beta$. We also impose a scalar weight $\gamma$ on the positive class (speaker active) to account for class imbalance when there are more speakers involved.

\vspace{-.2cm}
\subsection{Proposed visual-centric V-AHC}
\vspace{-.15cm}

Diarization in in-the-wild videos presents unique challenges, including speakers who are partially or completely off-screen and the presence of irrelevant on-screen speakers. However, visual signals such as face recordings are robust to acoustic noises, providing a strong cue about speaker identity and the places of articulation that discriminate between speech and non-speech signals. We aim to leverage the available visual signals, specifically the detected face tracks of on-screen speakers, to perform an on-screen speaker diarization. Our proposed visual-centric clustering-based model named V-AHC is illustrated in the upper half of Fig.~\ref{fig:av_model}, in the red dotted box.
\subsubsection{Speech activity extraction}
For every face track, we determine the speech activity by performing audio-visual active speaker detection using a pre-trained TalkNet model~\cite{tao2021someone}~\footnote{The TalkNet trained on the AVA-AVD dataset can be found at \url{https://github.com/TaoRuijie/TalkNet-ASD}}. It leverages the synchronization between the lip movements and the audio signals to determine whether there is an audible speech signal for the given face at the frame level. The use of audio signals is vital as it enhances the performance of visual signals as, in some circumstances, there is a talking face without audible speech.
\subsubsection{Speaker identity extraction}
To determine the identity of the speaker for a face track, we utilize deep face recognition models that have demonstrated remarkable performance in practical applications. Specifically, we randomly sample up to $50$ images from the face track and average their embeddings extracted from a pre-trained FaceNet model~\cite{schroff2015facenet}~\footnote{The pre-trained FaceNet~\cite{serengil2020lightface} can be found at \url{https://github.com/serengil/deepface}}.
\subsubsection{Agglomerative hierarchical clustering}
To identify speaker clusters, we utilize agglomerative hierarchical clustering~\cite{day1984efficient} (AHC) on all face tracks. The distances between face tracks are computed as the negative cosine similarity of their face embeddings averaged over 50 random frames. The diarization result for each speaker cluster is determined by combining the speech activities obtained with TalkNet for the respective face tracks. In cases where no face is detected for some segment, we set the corresponding speech activity to zero. 

\vspace{-.2cm}
\subsection{Permutation-invariant late fusion}
\vspace{-.15cm}
\label{sec:fusion}

The audio-only sub-system performs diarization on the entire recording, but is negatively impacted by acoustic noise. In contrast, the visual-centric sub-system is resilient to acoustic noise but will miss speech activities from off-screen speakers. To exploit the benefits of both the audio and visual models, we propose a score-level late-fusion strategy combining them by comparing their speech activity probabilities, as detailed next. As such, we aim to capitalize on their complementary strengths and enhance the overall diarization performance.

To determine the one-to-one correspondence between the audio and visual results, we compare every audio diarization result $\boldsymbol{\hat{y}}^s$ with every visual diarization result $\boldsymbol{\tilde{y}}^{s'}$ and calculate the matching score to be the summation of the audio scores $\hat{y}^s_t$ in time when visual scores $\tilde{y}^{s'}_t$ show active speech. The best correspondence is determined with the highest matching score for all permutations:
\begin{equation}     
\displaystyle{\displaystyle \argmax_{\pi \in \mathcal{P}_{(S,S')}} \displaystyle \sum_{s=1}^S \sum_{t=1}^T \hat{y}^{\pi(s)}_t \mathbb{1}[\tilde{y}^{\pi(s')}_t =1]} .
\end{equation}
For the best pairs, we replace the audio score $\hat{y}^s_t$ with the visual score $\tilde{y}^{\pi(s)}_t$ when the latter shows active speech, because we found the visual score to be more reliable. In some cases where we know the speaker overlapping ratio is small in the training data distribution, like AVA-AVD, we employ a post-processing technique that mutes the other speakers at those time frames where the visual score shows one speaker is active.

\vspace{-.2cm}
\section{EXPERIMENTAL SETUP}
\label{sec:experiment}
\vspace{-.15cm}

\subsection{Datasets}
\vspace{-.15cm}
\subsubsection{In the wild video: AVA-AVD}
The audio-visual AVA-AVD dataset~\cite{xu2021ava} is one of the few publically available in-the-wild video diarization datasets. It was built upon the AVA-Active Speaker dataset~\cite{roth2020ava}, which consists of multilingual movies depicting diverse daily activities, in order to foster the development of diarization methods for challenging conditions. The train, validation, and test sets consist of $243$, $54$, and $54$ videos respectively, $5$ minutes each. 

\subsubsection{Simulated proxy dataset: VoxCeleb2-AVD}
Since AVA-AVD is small, it is a standard practice to pre-train the models on a large simulated dataset.
We use the VoxCeleb2 dataset~\cite{Chung18b} (1 million YouTube videos from 6112 celebrities) as source material to simulate a pre-training dataset for the audio-only models, which we name VoxCeleb2-AVD. We simulate $2\!\times\!10^5$, $500$, and $500$ recordings for the train, validation, and test sets respectively, and each is $5$ minutes long. The test set has distinct speakers from the train and validation sets.

We strive to simulate VoxCeleb2-AVD as close as possible to AVA-AVD, as exemplified by the distributions %
shown in Fig.~\ref{fig:dataset}. For each audio recording, we first sample the number of speakers using a normal distribution with mean $8$ and standard deviation $2.5$, and the minimum (min) and maximum (max) number of speakers are set to 2 and 18, respectively. We select the speakers randomly from the dataset, then sample an utterance randomly from the selected speakers and concatenate all sampled utterances together. The length of each utterance is sampled from a truncated normal distribution with mean 0 and standard deviation $1.5$, and with min length of \SI{0.25}{\second}. With probability $0.8$, we add silence between the utterances, with the silence length sampled from a truncated normal distribution with mean $0.25$ and standard deviation \num{1}, and with min length of \SI{0.25}{\second}. With probability \num{0.2}, the following utterance is overlapped with the preceding utterance, with the overlapping duration uniformly sampled between \SI{0.25}{\second} and \SI{2}{\second}, and the max overlapping length is set to the length of the preceding utterance (i.e., we get at most $2$ overlapping speakers at any time).

We also randomly sample music and noise clips from the MUSAN dataset~\cite{musan2015} and the Freesound Dataset 50k (FSD50K)~\cite{fonseca2021fsd50k}, and add them to each audio recording. The noise clips are added such that they do not overlap with each other, and about 50\% of the length of each audio recording has noise added. The music clips are added similarly to the noise clips, but noise and music can overlap with each other. We follow the \textit{cocktail fork}~\cite{petermann2022cocktail} protocol in setting the energy levels between speech, noise, and music in VoxCeleb2-AVD, measuring them by loudness units full-scale (LUFS)~\cite{grimm2010toward}. The target LUFS values for speech, music, foreground noise, and background noise are set to $-17.0$, $-24.0$, $-21.0$, and $-29.0$ respectively. For each audio recording,  we first sample an average LUFS value for each class uniformly from a range of $\pm 2.0$ around the corresponding Target LUFS. Then each audio clip added to the audio recording has its individual gain further adjusted by uniformly sampling from a range of $\pm 1.0$ around the sampled recording-level LUFS value. We finally add \SI{0.1}{\second} fade-in and fade-out time for each audio clip.

\vspace{-.2cm}
\subsection{Implementation details}
\vspace{-.15cm}
We train the audio-only models on $5$-minute recordings. We sample the audio signals at 16 kHz, and obtain $40$-dimensional log-mel-filterbanks features with a $25$ millisecond (\SI{}{\milli\second}) frame length and \SI{10}{\milli\second} frame shift. Each feature is concatenated with those from the previous seven frames and the subsequent seven frames. We sub-sampled the concatenated features by a factor of ten, and the inputs for our model are thus $40 \!\times\! 15$ dimensional features every \SI{100}{\milli\second} similarly to~\cite{horiguchi2020end}. We used SA-EEND with $4$ encoder layers with hidden dimension $D$ set to $512$, feedforward dimension to $1024$, dropout to $0.1$, and number of heads to $8$. The encoder and decoder of the EDA are both $1$-layer LSTMs with a hidden size of $512$ and a dropout of $0.1$.

We set the scalar weights as $\alpha\,{=}\,0.01$ and $\beta\,{=}\,0.1\,{\times}\,0.92^{\text{\emph{Epoch}}}$, where \emph{Epoch} is the epoch number, as we empirically found that the speaker recognition loss is important during initial training iterations, but less beneficial as the model converges for diarization. We set $\gamma\,{=}\,5$. The model is trained using the Adam optimizer with the learning rate schedule proposed in~\cite{vaswani2017attention,horiguchi2020end} and \num{10000} warm-up steps. The batch size is set to $24$ on each \SI{48}{\giga\byte} RAM GPU using gradient accumulation. We train the model on $8$ GPUs, thus the effective batch size is $24\!\times\!8$. For both VoxCeleb2-AVD and AVA-AVD, we use the train set to train the model, the validation set to select the diarization threshold, and the test set to report the results.

The training of our proposed EEND-EDA++ requires the speaker identity labels in the dataset, and while VoxCeleb2 has such labels, AVA-AVD does not. Since the VoxCeleb2 training set has \num{5994} speakers, we can hope to find speakers in that dataset with similar voice characteristics as speakers in AVA-AVD. We thus map every speaker of AVA into its closest speaker in the VoxCeleb2 training set, based on the $L_2$ distance between their speaker embedding representations extracted using RawNet3~\cite{jung2022pushing}, as proxy for the speaker label.

\vspace{-.2cm}
\subsection{Baselines}
\vspace{-.15cm}

We present the results of $4$ baselines on the AVA-AVD dataset: WST~\cite{chung2019said}, VBx~\cite{landini2022bayesian}, AVR-Net~\cite{xu2021ava}, and DyViSE~\cite{dyvise2022}. WST is an audio-visual speaker diarization system that uses audio-visual correlation to help first enroll the speakers and then diarize. VBx is a state-of-the-art audio-only cascaded speaker diarization system that adopts Bayesian clustering; it has $2$ variants, VBx-ResNet34 and VBx-ResNet101, which are different in terms of the network layers extracting x-vectors~\cite{snyder2018x}. AVR-Net is an audio-visual cascaded speaker diarization system that is built upon VBx-ResNet34 and TalkNet~\cite{tao2021someone}. DyViSE is an audio-visual cascaded speaker diarization system, which denoises audio with visual information in a latent space and integrates facial features to obtain identity discriminative embeddings.

\vspace{-.2cm}
\section{RESULTS}
\label{sec:result}
\vspace{-.2cm}

\subsection{Audio-only models}
\vspace{-.15cm}
\subsubsection{Results on VoxCeleb2-AVD}
\vspace{-.15cm}

In Table~\ref{tab:voxceleb}, we present the results of the baseline EEND-EDA and our proposed EEND-EDA++ trained and evaluated on the simulated VoxCeleb2-AVD dataset. EEND-EDA++ achieves the best diarization error rate (DER) and Jaccard error rate (JER). Further decomposing DER, our method is better for all its submetrics: missed speech (MS), false alarm (FA), and speaker error (SE).

We also present two ablation studies of the proposed EEND-EDA++ on that same VoxCeleb2-AVD. EEND-EDA+Spk is trained with our speaker recognition loss, but without the attention mechanism. EEND-EDA+Spk performs badly in terms of DER and JER due to higher MS. This is probably because the vanilla EDA has limited capacity in producing attractors that are representative of speakers, thus the speaker loss adversely affects the model training. EEND-EDA+Att has the attention mechanism in EDA, but is not trained with our speaker recognition loss. EEND-EDA+Att outperforms EEND-EDA in DER, but is not better than EEND-EDA++ except for the MS submetric.

\begin{table}
    \centering
    \sisetup{
    detect-weight, 
    mode=text, 
    tight-spacing=true,
    round-mode=places,
    round-precision=1,
    table-format=2.1
    }
    \caption{Results on the simulated VoxCeleb2-AVD dataset. We assign a system number (Sys.) to each model. All systems in this table are only pre-trained on VoxCeleb2-AVD. We report the diarization error rate (DER), which is the sum of missed speech (MS), false alarm (FA), and speaker error (SE). We also report the Jaccard error rate (JER). The lower the better for all metrics. All results in this paper are with a collar of \SI{0.25}{\second}~\cite{istrate2005nist}.}
    \addtolength{\tabcolsep}{-.5pt}
    \resizebox{0.95\linewidth}{!}{
        \begin{tabular}{clSS[table-format=1.1]SSS} 
       \toprule
        Sys.    &Model  &\multicolumn{1}{c}{MS}     &\multicolumn{1}{c}{FA}     &\multicolumn{1}{c}{SE}   &\multicolumn{1}{c}{DER}    &\multicolumn{1}{c}{JER}    \\
        \midrule    
        7   &EEND-EDA~\cite{horiguchi2020end}
                        &17.4   &9.1    &18.9   &45.4   &66.7   \\
        \midrule
        9   &EEND-EDA++
                        &15.8	&6.7	&18.2	&40.8	&62.8   \\
        \midrule
        \midrule
        11  &EEND-EDA+Spk       &25.1   &6.4    &17.6   &49.1   &71.2   \\
        13  &EEND-EDA+Att       &14.6   &8.6    &20.8   &43.9   &66.6  \\
        \bottomrule
    \end{tabular}
    }
    \addtolength{\tabcolsep}{3.5pt}
    \vspace*{-4mm}
    \label{tab:voxceleb}
\end{table}

\vspace{-.2cm}
\subsubsection{Results on AVA-AVD}
\vspace{-.15cm}
In Table~\ref{tab:ava}, we compare our EEND-EDA++ with baselines on the AVA-AVD benchmark. The results of systems 1-4 are taken from~\cite{xu2021ava}; systems 2-4 use the same VAD, thus obtaining the same MS and FA. The previous SOTA is VBx-ResNet101 in system 3, which reports a DER of $70.9\%$. Without pre-training, EEND-EDA in system 6 performs badly with a DER of $96.4\%$. Systems 7-10 are pre-trained on our VoxCeleb2-AVD, and they all outperform the baselines by a wide margin in terms of DER, with our EEND-EDA++ achieving the best DER of $47.6\%$ and JER of $76.4\%$. It is worth mentioning that our EEND-EDA++ is an audio-only sub-system, but still outperforms by a wide margin the audio-visual baselines, that is, the WST, AVR-Net, and DyViSE models.

We also perform ablation studies for EEND-EDA+Spk and EEND-EDA+Att on the AVA-AVD benchmark. Similarly as for VoxCeleb2-AVD, %
EEND-EDA+Spk in systems 11 and 12 performs badly compared to the vanilla EEND-EDA model, mostly due to worse MS. EEND-EDA+Att also does not generalize well on AVA-AVD. Therefore, the speaker recognition loss and attention mechanism only help if used in tandem. It seems reasonable since the attention mechanism increases EDA's capacity in estimating more speaker attractors, 
while the speaker recognition loss regularizes the attention module.

\begin{table}
    \centering
    \sisetup{
    detect-weight, 
    mode=text, 
    tight-spacing=true,
    round-mode=places,
    round-precision=1,
    table-format=2.1
    }
    \caption{Results on the AVA-AVD benchmark. M indicates the modality used, either audio (A) or audio-visual (AV). PT indicates if the model is pre-trained, either on a VoxCeleb2-based dataset (Sys.~1-5) as in~\cite{xu2021ava}, on our VoxCeleb2-AVD (Sys.~7-16 and 18-21), or on the AVA and some face recognition datasets (Sys.~17). FT indicates if the model is fine-tuned on AVA-AVD. *Systems 15 and 16 use the ground-truth number of speakers at inference.}
    \addtolength{\tabcolsep}{-3.5pt}
    \resizebox{\linewidth}{!}{
    \begin{tabular}{clcccSSSSS} 
       \toprule
        Sys.    &Model  &M &PT     &FT     &\multicolumn{1}{c}{MS}     &\multicolumn{1}{c}{FA}     &\multicolumn{1}{c}{SE}   &\multicolumn{1}{c}{DER}    &\multicolumn{1}{c}{JER}    \\
        \midrule    
        1   &WST~\cite{chung2019said}  &AV      
                        &\multirow{5}{*}{\cmark}
                                            &\multirow{5}{*}{\cmark}
                                                    &11.6   &40.6   &36.1   &88.4   &\multicolumn{1}{c}{-}      \\
        2   &VBx-ResNet34~\cite{landini2022bayesian} &A
                        &        &       &8.7    &44.6   &35.3   &88.5   &\multicolumn{1}{c}{-}      \\
        3   &VBx-ResNet101~\cite{landini2022bayesian} &A
                        &        &       &8.7    &44.6   &17.6   &70.9   &\multicolumn{1}{c}{-}      \\
        4   &AVR-Net~\cite{xu2021ava} &AV
                        &        &       &8.7    &44.6   &20.1   &73.3   &\multicolumn{1}{c}{-}      \\
        5   &DyViSE~\cite{dyvise2022} &AV
                        &        &       &11.1   &24.2   &35.93   &71.2   &\multicolumn{1}{c}{-}      \\
        \midrule
        6   &EEND-EDA~\cite{horiguchi2020end} &\multirow{3}{*}{A}
                        &\xmark &\cmark &46.6   &27.3   &22.6   &96.4   &94.7   \\
        7   &EEND-EDA &
                        &\cmark &\xmark &24.2   &2.8    &24.4   &51.4   &83.5   \\
        8   &EEND-EDA &
                        &\cmark &\cmark &28.5   &5.3    &15.1   &48.9   &78.5   \\  
        \midrule
        9   &EEND-EDA++ &\multirow{2}{*}{A}
                        &\multirow{2}{*}{\cmark} 
                                &\xmark &22.8   &6.8    &20.8   &50.4   &80.5  \\
        10  &EEND-EDA++ &
                        &       &\cmark &25.0   &5.6    &17.0   &47.6   &76.4  \\ 
        \midrule
        \midrule
        11  &EEND-EDA+Spk   &\multirow{4}{*}{A} &\multirow{4}{*}{\cmark}
                                    &\xmark &29.8   &4.0    &21.8   &55.6   &84.5   \\
        12  &EEND-EDA+Spk   &&       &\cmark &35.4   &3.3    &16.4   &55.1   &82.7   \\
        13  &EEND-EDA+Att   &&       &\xmark &28.3   &3.6    &20.1   &52.1   &82.2   \\
        14  &EEND-EDA+Att   &&       &\cmark &38.2   &1.8    &14.2   &54.2   &82.0   \\
        \midrule
        15  &EEND-EDA*   &\multirow{2}{*}{A}
                        &\multirow{2}{*}{\cmark}  &\multirow{2}{*}{\cmark}  
                        &28.7	&4.7	&15.3   &48.7    &78.5 \\  
        16  &EEND-EDA++* &&&
                        &23.2	&7.8	&16.7   &47.7    &74.3 \\
        \midrule
        \midrule
        17  &V-AHC     &\multirow{5}{*}{AV} &\multirow{5}{*}{\cmark} &\multirow{5}{*}{\cmark}
                        &56.2   &0.9    &9.2    &66.3   &78.4 \\
        18  &AV-EEND-EDA    &&&
                        &20.4   &5.9    &19.1   &45.4   &70.5 \\
        19  &AV-EEND-EDA++    &&& 
                        &18.5   &7.2    &20.4	&46.1	&68.8\\
        20  &AV-EEND-EDA$^\dagger$    &&&
                        &21.3	&3.9	&19.8	&45.0	&77.3 \\
        21  &AV-EEND-EDA++$^\dagger$    &&&
                        &20.0	&4.66	&20.4	&45.1	&76.0\\
        \bottomrule
    \end{tabular}
    }
    \addtolength{\tabcolsep}{3.5pt}
    \vspace*{-6mm}
    \label{tab:ava}
\end{table}


\vspace{-.2cm}
\subsubsection{Results on AVA-AVD with oracle speaker counting}
\vspace{-.15cm}

We also present systems 15 and 16 in Table~\ref{tab:ava} 
to study the effects of the estimation of the number of speakers in diarization performance. For baseline EEND-EDA, system 15 is the same as system 8, except that the former uses the ground-truth number of speakers at inference. %
Both get similar DER and JER, but with different MS/FA trade-offs. %
For EEND-EDA++, system 16 is the same as system 10, except that the former uses the ground-truth number of speakers at inference. Both get similar DER, but system 10 is lagging behind system 16 by $2\%$ in terms of JER. We observe informally that system 10 typically underestimates the number of speakers. Nevertheless, system 10 still achieves the best DER and JER among non-oracle systems 1-14, credited to our better attractor representation. 

\vspace{-.2cm}
\subsubsection{Visualization}
\begin{figure}[t]
\begin{minipage}[t]{\linewidth}
  \centering
  \includegraphics[width=0.7\linewidth]{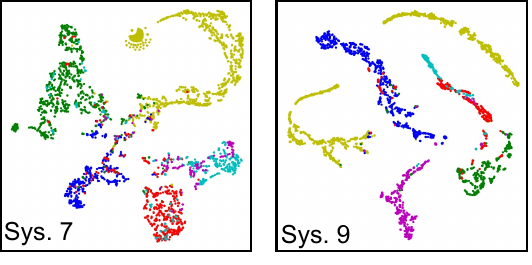}
  \vspace*{-4mm}
  \caption{t-SNE plot of VoxCeleb2-AVD embeddings $e_t$ in a recording}\medskip
  \label{fig:vox_embedding}
\end{minipage}
\hfill
\begin{minipage}[t]{\linewidth}
  \centering
  \centerline{\includegraphics[width=0.7\linewidth]{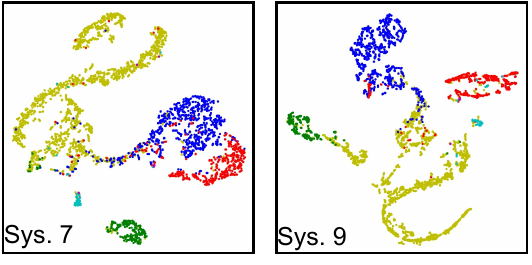}}
  \vspace*{-4mm}
  \caption{t-SNE plot of AVA-AVD embeddings $e_t$ in a recording}\medskip
  \label{fig:ava_embedding}
\end{minipage}
\hfill
\begin{minipage}[t]{\linewidth}
  \centering
  \centerline{\includegraphics[width=0.7\linewidth]{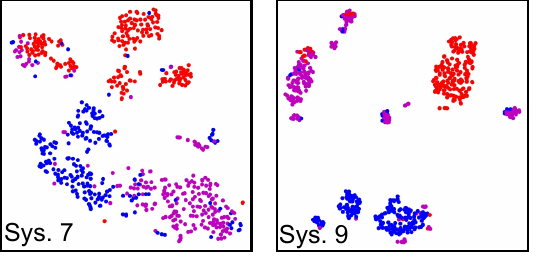}}
  \vspace*{-4mm}
  \caption{t-SNE plot of VoxCeleb2-AVD attractors $a_s$ across recordings}\medskip
  \label{fig:vox_attractor}
\end{minipage}
\vspace*{-6mm}
\end{figure}

\vspace{-.15cm}
In Figs.~\ref{fig:vox_embedding} and~\ref{fig:ava_embedding}, we show the t-SNE plot of embeddings $e_t$ within one recording for the VoxCeleb2-AVD and the AVA-AVD dataset. Yellow represents the non-speech region, while each of the other colors represents a speaker. For both datasets, the embedding clusters of our system 9 are separated further apart than the baseline system 7, showing that our speaker-recognition objective pushes the embeddings from different speakers away from each other.

In Fig.~\ref{fig:vox_attractor}, we show the t-SNE plot of attractors across different recordings from the VoxCeleb2-AVD dataset. We randomly selected $3$ speakers who are not seen during training, and if these speakers appear in a recording, we match the attractors to the speaker labels by computing the best permutation between the estimated diarization streams and ground-truth diarization streams. We then plot the attractors that were matched to the three speakers. Each color represents a speaker. We see that our system 9 has obvious clusters for the attractors compared to baseline system 7, which means that the speaker identities are better matched across recordings. Note we cannot show attractor plots for AVA-AVD as it does not have ground-truth speaker labels.

\vspace{-.2cm}
 \subsection{Visual-centric model}
\vspace{-.15cm}
In Table~\ref{tab:ava}, our proposed visual-centric on-screen speaker diarization sub-system V-AHC achieves a DER of $66.3\%$, which surpasses the baseline systems 1 to 5. V-AHC has a very high MS of $56.2\%$, which shows that the off-screen speakers are indeed a frequent feature in the AVA-AVD dataset.  
However, V-AHC has very low FA of $0.9\%$, showing the strength of the visual signals when present.




\subsection{Audio-visual models}
We fuse the visual-centric sub-system V-AHC with the audio-only sub-systems 8 and 10, resulting in the audio-visual systems 18 and 19 respectively.
We can see that the fused audio-visual models outperform their audio-only sub-systems respectively, system 18 outperforming system 8 with an absolute reduction of $3.5\%$ in DER and $8\%$ in JER, and system 19 outperforming system 10 with an absolute reduction of $1.5\%$ in DER and $7.6\%$ in JER. 
Compared to audio sub-system counterparts, the MS generally decreases significantly with the fusion model, but the FA and SE increase. This may be because the final number of speakers is set as the maximum of the audio and visual sub-systems, with an increase in the number of output streams, the MS decreases, but introducing more FS and SE. 

We also present AV-EEND-EDA$^\dagger$ (sys. 20) and AV-EEND-EDA++$^\dagger$ (sys. 21), in which instead of fusing the full-length recording-level visual diarization results to the audio diarization results as described in Sec~\ref{sec:fusion}, we fuse each face activities output from the TalkNet (face-track-level) to one of the audio diarization streams, with the same score-level decisions. 
System 20 outperforms system 8 with an absolute reduction of $3.9\%$ in DER and $1.2\%$ in JER, and system 21 outperforms system 10 with an absolute reduction of $2.5\%$ in DER and $0.4\%$ in JER. 
Compared to audio sub-system counterparts, the MS and FA generally decrease, but SE increases. This is because the final number of output speakers is the same as the audio sub-system, so the visual signal generally helps the MS and FA. However, the fusion without considering visual speaker identity causes the SE to increase.
Although the improvements on DER here are better compared to recording-level audio-visual fusion, the JER here only improves marginally. Overall recording-level fusion is preferred as the improvement on JER is large which represents per speaker diarization evaluation improves a lot, showing the importance of utilizing the identity information from the visual signals by the FaceNet and the AHC algorithm.

\vspace{-.2cm}
\section{CONCLUSION}
\label{sec:conclusion}
\vspace{-.15cm}

We studied the speaker diarization problem for in-the-wild videos. We proposed a late audio-visual fusion model, AV-EEND-EDA++, that comprises an audio-only sub-system, EEND-EDA++, and a visual-centric sub-system, V-AHC. For audio-only sub-systems, we show that E2E audio-only systems demonstrate remarkable performance if it is trained on our proposed dataset recipe, and the speaker identity are better preserved in our EEND-EDA++ by using our speaker recognition loss. Our proposed sub-systems and the fused audio-visual model outperform SOTA on the AVA-AVD benchmark.




\balance
\bibliographystyle{IEEEbib}
\bibliography{refs}

\end{document}